# Whistler Wave Observations by *Parker Solar Probe* During Encounter 1: Counter-Propagating Whistlers Collocated with Magnetic Field Inhomogeneities and their Application to Electric Field Measurement Calibration

S. Karbashewski [1] O. V. Agapitov [1,2] H. Y. Kim [1] F. S. Mozer [1] J. W. Bonnell [1] C. Froment [3] T. Dudok de Wit [3,4] Stuart D. Bale [5,1] D. Malaspina [1] and N. E. Raouafi [6]

[1] Space Sciences Laboratory, University of California, Berkeley, CA 94720, USA
[2] Astronomy and Space Physics Department, National Taras Shevchenko University of Kyiv, Kyiv, Ukraine
[3] LPC2E, CNRS, CNES, University of Orléans, Orléans, France
[4] International Space Sciences Institute, Bern, Switzerland
[5] Physics Department, University of California, Berkeley, CA 94720-7300, USA
[6] Johns Hopkins University Applied Physics Laboratory, Laurel, MD 20723, USA

## ABSTRACT

Observations of the young solar wind by the Parker Solar Probe (PSP) mission reveal the existence of intense plasma wave bursts with frequencies between 0.05 — $0.20 f_{\rm ce}$ (tens of Hz up to ∼300 Hz) in the spacecraft frame. The wave bursts are often collocated with inhomogeneities in the solar wind magnetic field, such as local dips in magnitude or sudden directional changes. The observed waves are identified as electromagnetic whistler waves that propagate either sunward, anti-sunward, or in counter-propagating configurations during different burst events. Being generated in the solar wind flow the waves experience significant Doppler down-shift and up-shift of wave frequency in the spacecraft frame for sunward and anti-sunward waves, respectively. Their peak amplitudes can be larger than 2 nT, where such values represent up to 10% of the background magnetic field during the interval of study. The amplitude is maximum for propagation parallel to the background magnetic field. We (i) evaluate the properties of these waves by reconstructing their parameters in the plasma frame, (ii) estimate the effective length of the PSP electric field antennas at whistler frequencies, and (iii) discuss the generation mechanism of these waves.

## 1. INTRODUCTION

NASA's Parker Solar Probe (PSP), launched in August 2018, is the first mission to collect *in situ* measurements of solar wind parameters within the orbit of Mercury (Fox et al. 2016). Following an initial gravity assist with Venus, PSP was inserted into its first encounter with the Sun and reached a perihelion of $35.67 R_\odot$ (∼0.166 AU) from the Sun in November of 2018. Through a subsequent series of six gravity assists with Venus between 2019 and 2024, the perihelion of PSP's orbit will progressively shrink through 24 encounters to $9.86 R_\odot$ (∼0.046 AU) (Fox et al. 2016). The mission is designed to address three core science objectives: (i) trace the flow of energy that heats the solar corona and accelerates the solar wind, (ii) determine the structure and dynamics of the plasma and magnetic fields at the sources of the solar wind, and (iii) explore mechanisms that accelerate and transport energetic particles. Wave-particle interactions involving both magnetohydrodynamic (MHD) and kinetic-scale waves, such as Alfvén and whistler waves, respectively, are ubiquitous in the solar wind and known to play an important part in addressing all three science objectives of the mission. In particular, wave-particle interactions are central to understanding the evolution of the core, halo, and strahl suprathermal electron populations (Feldman et al. 1975).

Corresponding author: S. Karbashewski
sgkarbo@berkeley.edu

The first encounter phase of PSP lasted from October 31, 2018, to November 12, 2018; during which the near co-rotation of the probe with an equatorial coronal hole meant the probe was immersed in a slow (less than 500 km/s) and highly Alfvénic solar wind (Bale et al. 2019; Kasper et al. 2019). Halekas et al. (2020) reported that during the encounter the density and temperature of the core electron population increased with heliocen-tric distance and agreed with previous empirical radial scaling. The strahl population was more narrowly distributed at perihelion while the halo population nearly disappeared – similar to reports at 0.3 AU by Štverák et al. (2009); given this, the strahl population dominated the suprathermal fraction by nearly a factor of 10 over the halo population at the closest approach. The observed fractional halo population observed (less than 1%) is significantly less than that reported between 1 AU to 4 AU where it typically represents between 4% and 10% of the total population depending on the fitting procedures used (McComas et al. 1992; Štverák et al. 2009).

The evolution of the suprathermal populations is such that the halo increases with distance from the Sun while the strahl population decreases (Maksimovic et al. 2005; Štverák et al. 2009). This naturally leads to the interpretation of an exchange between the populations, with strahl electrons transferred to the halo through scattering in wave-particle interactions. There is strong evidence that whistler waves are, at least partially, responsible for this conversion process. Whistler waves are known to strongly interact with high energy electron populations and have been observed to scatter the strahl in the solar wind at 1 AU (Kajdič et al. 2016) and control the dynamics of the relativistic electron populations in the Earth's radiation belts (Horne 2007; Thorne 2010).

There was strong whistler wave activity throughout much of PSP's Encounter 1 and several authors have reported on these observations (Agapitov et al. 2020; Dudok de Wit et al. 2022; Cattell et al. 2021b,a, 2022; Froment et al. 2022). Recently, whistler waves observed during Encounter 1 have been linked to the scattering of strahl at distances less than $50R_{\odot}$ (less than 0.25 AU) from the Sun (Jagarlamudi et al. 2021; Cattell et al. 2021a,b) including some of the cases reported in this article. The statistical study by Cattell et al. (2021b) of whistlers observed by PSP concludes that the waves are more intermittent closer to the Sun (less than 0.3 AU) than at 1 AU, have frequencies on the order of approximately $0.2f_{\rm ce}$ and are often associated with higher magnetic activity. These periods of whistler wave activity are interspersed with higher frequency electron Bernstein modes in the range of $f_{ce}$ and its harmonics Malaspina et al. (2020). Recent work has demonstrated that the observed electron Bernstein modes are most likely generated by the interaction between the solar wind and the Parker Solar Probe spacecraft wake Malaspina et al. (2022); Tigik et al. (2022). Nonetheless, electron Bernstein waves are found to preferentially occur in regions of exceptionally low-amplitude magnetic turbulence, termed quiescent regions Short et al. (2022). The lack of whistler observations in quiescent regions is consistent with the connection between whistler activity and magnetic structures reported in the current study.

The local magnetic field has a strong impact on the propagation of whistler-mode waves, and any inhomogeneities in the field strength or direction can lead to the reflection, refraction, absorption, or mode conversion of such waves. An interesting feature of the magnetic field of the solar wind measured during the PSP encounters is the observation of so-called "magnetic switchbacks" (Krasnoselskikh et al. 2020; Agapitov et al. 2022); these are defined as rapid deflections of the magnetic field that last seconds to hours before "flipping" back to the original field state and are sometimes observed as complete reversals of the magnetic field direction. A feature of the leading and trailing boundaries of the switchback structures is the presence of dips in the magnitude of the magnetic field (by a few percent) and enhanced wave activity, including whistler waves (Krasnoselskikh et al. 2020). A recent discovery made by PSP and reported in (Agapitov et al. 2020) is the existence of sunward propagating whistler waves collocated with these magnetic dips. The waves represented fluctuation amplitudes up to 10% of the background magnetic field. The sunward propagating feature makes these waves a prospective candidate for the efficient scattering of strahl electrons into the halo population (Saito & Gary 2007).

In this article, we focus on the observation of whistler waves by PSP at $\sim 42.6R_{\odot}$ ($\sim$0.2 AU) that are collocated with magnetic inhomogeneities (changes in magnetic field magnitude or magnetic field direction on the time scale of 1-2 s), including the edges of switchback structures, but also at local rotations of the magnetic field and magnetic dips not associated with switchbacks. We focus on how the relative (in the solar wind frame) velocity of the inhomogeneities, which are presumably the source of the whistler waves, determines their spectral properties and propagation direction. We consider five localized whistler wave packets captured during a one-hour interval in the high sampling frequency "burst mode" of PSP. These cases have similar plasma conditions, such as density, magnetic field magnitude, suprathermal electron population, and electron distribution anisotropy. The whistler waves are observed to propagate predominantly radially in counter-propagating, sunward, and anti-sunward directions. We determine the fine spectral-temporal structure of these waves in the plasma frame by using the high-frequency burst magnetic and electric fields data and utilize these observations, along with the cold homogeneous whistler-mode dispersion relation, to estimate the effective length of the electric field antennas that are part of the PSP FIELDS instrument.

The paper is organized as follows: Section 2 contains the data description and the analysis of PSP data highlighting bursts of counter-propagating whistler waves; Section 3 is an application of the whistler wave measurements to a determination of the effective length of the electric field antennas that are part of the FIELDS instrument; and Section 4 is a discussion of the results and proposed wave generation mechanism and highlights the main conclusions.

## 2. PARKER SOLAR PROBE OBSERVATIONS

In this section, we investigate whistler waves in the solar wind using measurements from PSP on November 3, 2018, during Encounter 1 at the heliocentric distance of $40R_{\odot}$. The coordinate system used throughout this manuscript is the inertial "RTN" (Radial-Tangential-Normal) system with the radial component R oriented along the Sun-spacecraft line, the transverse component T is defined to be orthogonal to the rotational axis of the Sun and the radial component, i.e. $\mathrm{T} = \Omega_{\odot} \times \mathrm{R}$, while the normal component N completes the orthogonal right-handed triad and, in this case, is aligned with the normal of the ecliptic plane (Russell et al. 2016). The electromagnetic fields are measured by the FIELDS instrument suite on PSP (Bale et al. 2016). The electric field measurements are made using the electric fields instrument (EFI) consisting of two pairs of dipole electric field antennas oriented in the TN-plane and extending beyond the PSP heat shield, and a fifth antenna located behind the heat shield on the instrument boom; the location of antenna $V5$ in the wake of PSP means the R-component is susceptible to detrimental interference by the wake electric field and cannot be reliably interpreted (Bale et al. 2016). Two, three-component flux gate magnetometers (MAG) measure the magnetic field from DC to approximately 60 Hz (up to 293 measurements per second during 2-4 days around perihelion) (Bale et al. 2016), while the three-component Search Coil Magnetometer (SCM) measures the magnetic field perturbations above approximately 3 Hz up to approximately 1 MHz (Dudok de Wit et al. 2022; Jannet et al. 2021). The Digital Fields Board (DFB) is responsible for the signal processing and digitization of these signals and outputs a variety of data products for analysis (Bale et al. 2016; Malaspina et al. 2016). The DFB Burst Memory (DBM) data product is an output of approximately 3.5 s of waveform data sampled at 150 kS/s, providing more than adequate coverage of the whistler frequency range that was typically below 1 kHz during the first encounter. The SWEAP instrument suite on PSP is responsible for measurements of the solar wind protons, $\alpha$ particles, and electrons including their densities, temperatures, and bulk velocities (Kasper et al. 2016). In the present analysis, we focus primarily on DBM measurements of the electric and magnetic fields using the EFI and SCM, but we also make use of the background magnetic field measurements from the MAG, low cadence spectrograms of the EFI and SCM covering the whistler frequency range, and we use the Solar Probe Cup (SPC) instrument on SWEAP to derive the proton density and velocity (Case et al. 2020).

The observations investigated here occurred on November 3, 2018, between the hours of 10:00 and 11:00 UTC. Fig. 1(a) presents the solar wind magnetic field dynamics recorded by the MAG during this observation window. The magnitude of the background magnetic field is indicated by the black curves and the three RTN components by the red, blue, and green curves. The magnitude of the interplanetary magnetic field varies between 40 nT and 70 nT and the radial component of the magnetic field vector is primarily directed towards the Sun (negative R direction). This time period is characterized by a high level of magnetic field perturbations and we observe switchback-like structures, rapid rotational discontinuities, and magnetic dips of varying magnitudes and duration.

Figs. 1(b) and (c) present the frequency spectra captured by the SCM and EFI, respectively. The magnetic field spectra are represented by a quadrature summation of the three components while the electric spectra are only available at this time, in this frequency range, for the differential voltage signal between antennas 1 and 2, oriented in the TN plane. The upper black curve represents $0.1f_{\mathrm{ce}}$ and the lower black curve is the local lower hybrid frequency $f_{\mathrm{lh}}$, which is roughly the lower limit of the whistler frequency range. There is extensive wave activity in the lower (presumably magneto-hydrodynamic, MHD) frequency range corresponding predominantly to Alfvénic fluctuations and in the whistler frequency range from approximately 35 Hz to 500 Hz ($0.02 — 0.3f_{\mathrm{ce}}$). We note that the large bursts of whistler waves are often collocated with periods of high magnetic activity corresponding to magnetic field inhomogeneities like magnetic field magnitude local minima or rapid magnetic field direction changes; this correlation will be discussed further in Section 4.

The solar wind proton velocity determined by the SPC instrument is shown in Fig. 1(e). The R-component of solar wind velocity dominates with a typical magnitude in the range of 250 km/s to 320 km/s. The variation in the structure of the solar wind velocity for each of the coordinates is highly correlated with the corre-sponding magnetic component (switchbacks, dips, rotations), however, while the radial magnetic field often approaches zero and sometimes even flips sign the radial velocity component maintains a large magnitude and is always flowing out from the Sun. The density from SPC shown in Fig. 1(f) varies between roughly 250 cm$^{-3}$ and 400 cm$^{-3}$ and the structural variation is similarly seen be correlated with the magnetic activity.

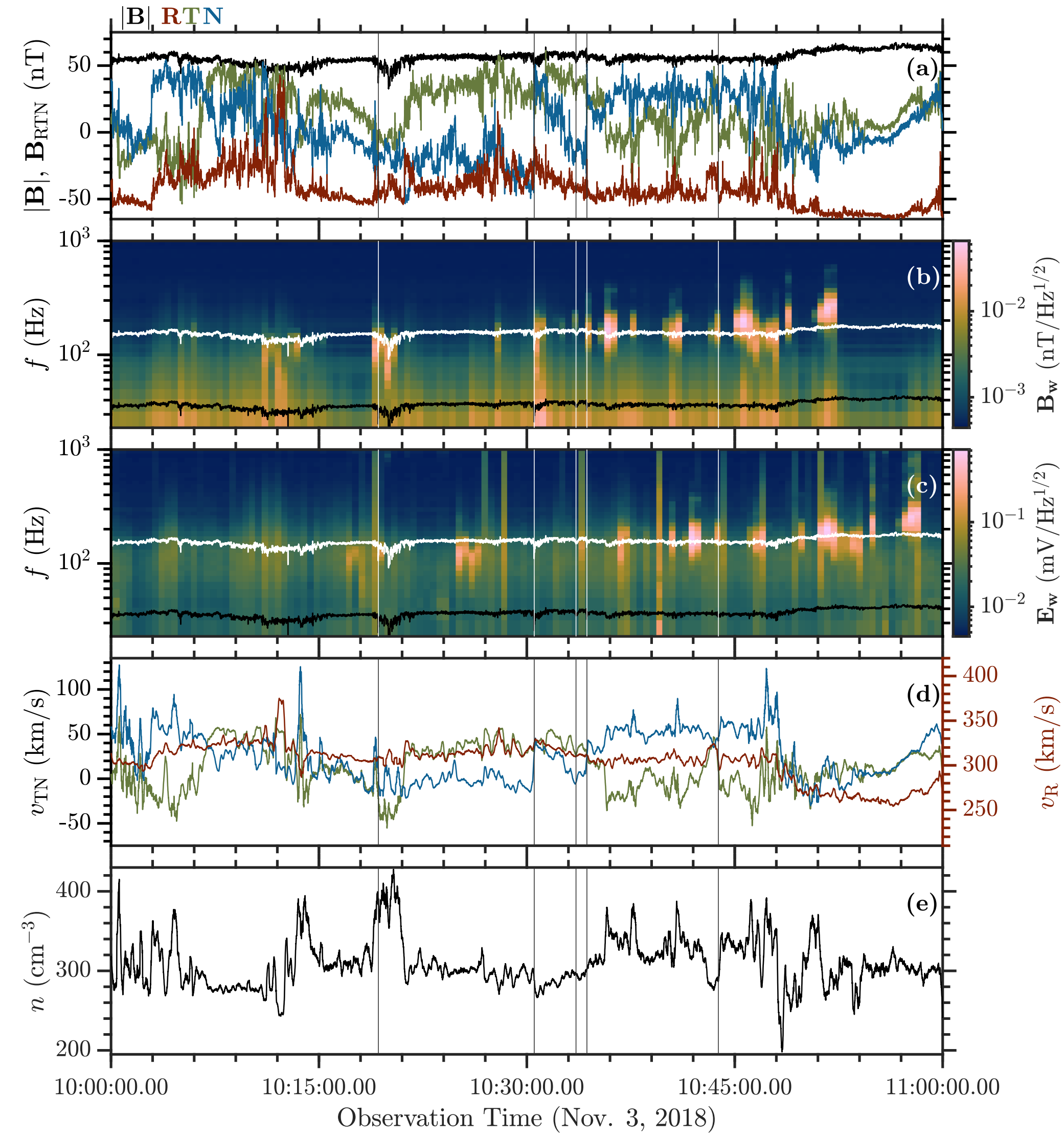

**Figure 1.** Solar wind parameters during the one-hour interval where the whistlers were observed on November 3, 2018, between 10:00 and 11:00 UTC. (a) The magnetic field (RTN coordinates in red, green, and blue, respectively, and the magnetic field magnitude is in black) from the MAG. (b) SCM magnetic field spectra, the black line indicates the lower hybrid frequency, $f_{\rm lh}$, and the white line is $0.1 f_{\rm ce}$. (c) DC coupled electric field spectra from the potentials difference of antennas 1 and 2. (d) Solar wind bulk flow velocity in RTN coordinates (same colors as (a) with $v_{\rm TN}$ on the left axis and $v_{\rm R}$ on the right axis) from SWEAP/SPC. (e) Plasma proton density obtained from SWEAP/SPC. The vertical lines in all panels indicate the location of burst waveform intervals from the DBM.

The DBM sampling intervals during this one-hour interval are highlighted by the 5 vertical lines between 10:15:00 UTC and 10:45:00 UTC. In this particular observation interval, all of these 3.5 s high sampling rate periods coincided with significant whistler wave activity, with four of them collocated with magnetic dips and/or rotations. We use three components of the magnetic field transferred into the RTN coordinate system and two components of the electric field recorded in the spacecraft plane transverse to the direction of the Sun (the $R$-axis of the RTN coordinate system) transferred into $T$ and $N$ components. The third $R$-component of the electric field is calculated in the whistler frequency range from $\vec{E}_{\rm w} \cdot \vec{B}^*_{\rm w} = 0$, where $\vec{E}_{\rm w}$ and $\vec{B}_{\rm w}$ are spectral components of the magnetic and electric field. This

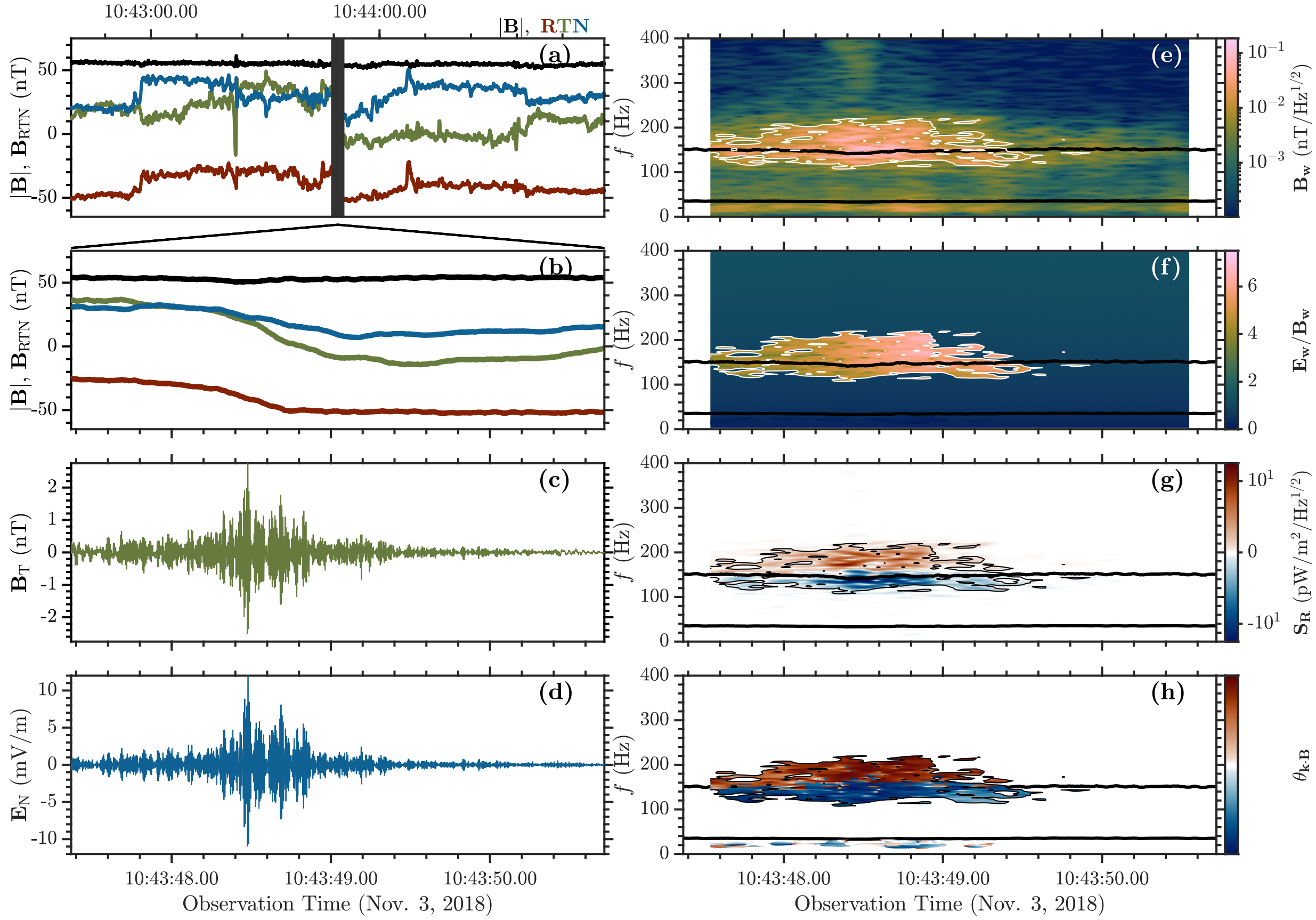

**Figure 2.** The burst interval collected staring at 10:43:47.29 UTC on November 3, 2018. (a) Dynamics of the magnetic field in RTN coordinates from the MAG over a short time interval around the burst. (b) The shaded region of (a) shows the magnetic field only during the burst. (c) Burst waveform of the magnetic field T-component, $\delta B_{\mathrm{T}}$. (d) Burst waveform of the electric field N-component, $\delta E_{\mathrm{N}}$. (e) Spectrogram of the magnetic field burst waveforms (the white contour indicates the spectral power threshold applied to the data in panels (f-h)). (f) The ratio of electric spectrogram to magnetic wave spectral power; the white contour denotes the data boundary with lower amplitude regions omitted for clarity and the background coloring represents the expected $E_{\mathrm{w}}/B_{\mathrm{w}}$ ratio for parallel propagating whistler waves. (g) The radial component of the Poynting vector, $S_R$, is calculated from the spectrograms of each component and transformed to a symmetric log scale as discussed in the text. (h) Wave normal angle with respect to the background magnetic field, $\theta_{\mathbf{k}\cdot\mathbf{B}}$, ranging from 0° to 180° and indicating parallel and anti-parallel propagation, respectively. The lower and upper solid black curves in (e)-(h) indicate $f_{\mathrm{lh}}$ and $0.1 f_{\mathrm{ce}}$, respectively.

gives for the $E_{\mathrm{w}R}$:

$$E_{\mathrm{w}R} = (E_{\mathrm{w}T} \cdot B^*_{\mathrm{w}T} + E_{\mathrm{w}N} \cdot B^*_{\mathrm{w}N}) * B_{\mathrm{w}R}/|B_{\mathrm{w}R}|^2. \quad (1)$$

This estimation of $E_{\mathrm{w}R}$ is used to process the spectral power of the electric field in the processing of the $E_{\mathrm{w}}/B_{\mathrm{w}}$ ratio and evaluating the effective length of the PSP electric field antennas in Section 3.

We first turn our attention to the case that begins at 10:43:47.29 UTC. Fig. 2(a) shows the magnetic field captured by the MAG for a few minutes surrounding the event with the duration of the actual burst event highlighted by the gray-shaded region. The wave burst is collocated with a magnetic rotation associated with the trailing edge of a larger magnetic event that has the typical characterizations of a magnetic switchback (observed from approximately 10:43:00 UTC to 10:43:48 UTC). The magnetic rotation also results in a depletion of the magnetic field magnitude by about 10% from the background field of approximately 55 nT. Fig. 2(b) displays the MAG measurement during the DBM event highlighted by the gray shading in Fig. 2(a); the rapid rotation of the magnetic field, taking less than one second, is more clearly seen.

Figs. 2(c) and (d) present the T-component of the DBM magnetic and N-component of the DBM electric field measurements recorded by the SCM and EFI, respectively. The wave packet, which will be shown to be in the whistler frequency range, reaches an amplitude greater than 2 nT peaking on the maximum of the gradient of the magnetic rotation and is reduced significantly towards the end of the DBM event. The frequency spectrogram of the three-component magnetic field DBM, $B_w$, is shown in Fig. 2(e) with the upper and lower curves representing $0.1f_{\rm ce}$ and $f_{\rm lh}$ as in Fig. 1. The bulk of the wave power is contained between 100 Hz and 250 Hz and belongs to the whistler wave frequency range; this leaves little doubt that these are indeed electromagnetic whistler waves. The ratio $E_{\rm w}/B_{\rm w}$ displayed in Fig. 2(f) in the spacecraft frame is calculated using the 3-component magnetic spectra and 3-component electric spectra (with the $R$-component estimated from $\vec{E}\cdot\vec{B}=0$). The white contour border outlines the regions of high wave spectral density from Fig. 2(e) with the rest of the data omitted; the background color indicates the $E_{\rm w}/B_{\rm w}$ ratio expected for parallel propagating whistlers at each frequency. Importantly, the ratio is significantly larger than the expectation for whistler waves, as has been observed by others (Agapitov et al. 2020); this outcome is a by-product of the EFI antennas having a frequency-dependent effective length and will be addressed in Section 3.

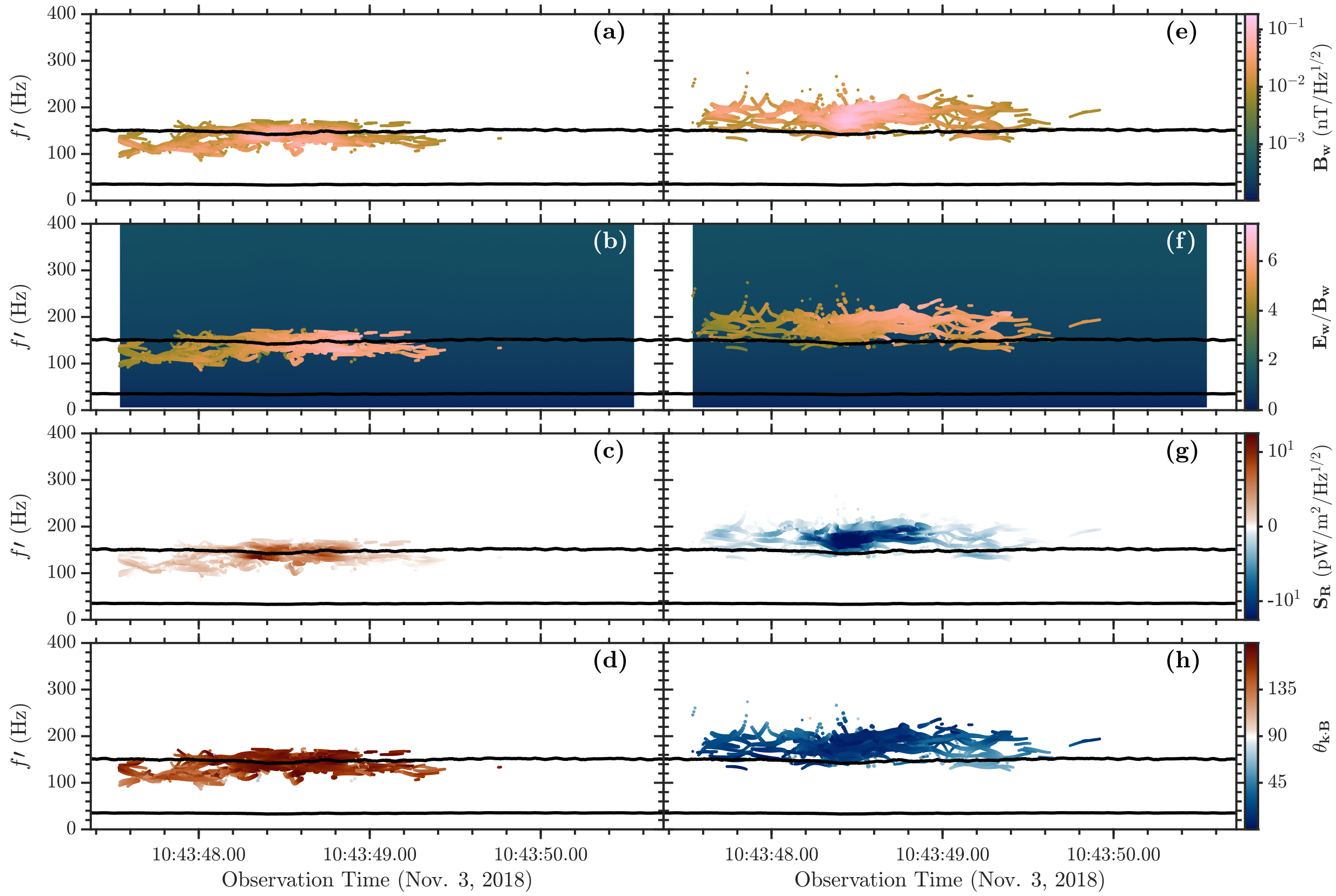

**Figure 3.** Doppler shifted frequencies, $f\prime$, of the burst waveform spectrogram starting at 10:43:47.29 UTC on November 3, 2018; only the largest amplitude whistles (above the threshold in Figure 2e) are retained. (a/e) The magnitude of magnetic field fluctuations. (b/f) $E_{\rm w}/B_{\rm w}$ ratio with the expected ratio represented by the colored background. (c/g) R-component of the Poynting vector. (d/h) Wave normal angle with respect to the background magnetic field. The left column (a-d) is for the anti-sunward propagating whistlers and the right column (e-h) is for the sunward propagating whistlers. The upper and lower solid black curves indicate $0.1f_{\rm ce}$ and $f_{\rm lh}$, respectively.

The Poynting flux in the spacecraft frame is calculated using the complex $\vec{E}_{\rm w}$ and $\vec{B}_{\rm w}$ spectra shown in Fig. 2(g) (Webber 2012). The most striking feature of this analysis is the existence of both sunward and anti-sunward waves, which we have elected to call counter-propagating whistler waves. The counter-propagating nature is profoundly strong evidence that the PSP spacecraft is traveling through, or very near to, the source region of the whistler waves; such a burst of sunward and anti-sunward waves occurring simultaneously with the symmetry observed here can be presumably explained by this scenario. Fig. 2(h) shows the wave normal angle (WNA) between the wave vector and the background magnetic field, $\theta_{\mathbf{k}\cdot\mathbf{B}}$ (the R-component of the background magnetic field from the MAG was negative during the event, i.e. directed toward the Sun) using the magnetic spectral matrices calculated from three-component SCM measurements (Santolík et al. 2003); this reinforces the counter-propagating feature and reveals the waves are propagating mostly quasi-parallel ($< 20°$) to the magnetic field but do reach oblique angles of propagation up to 60° (but with much lower amplitude) because of curvature of the background magnetic field.

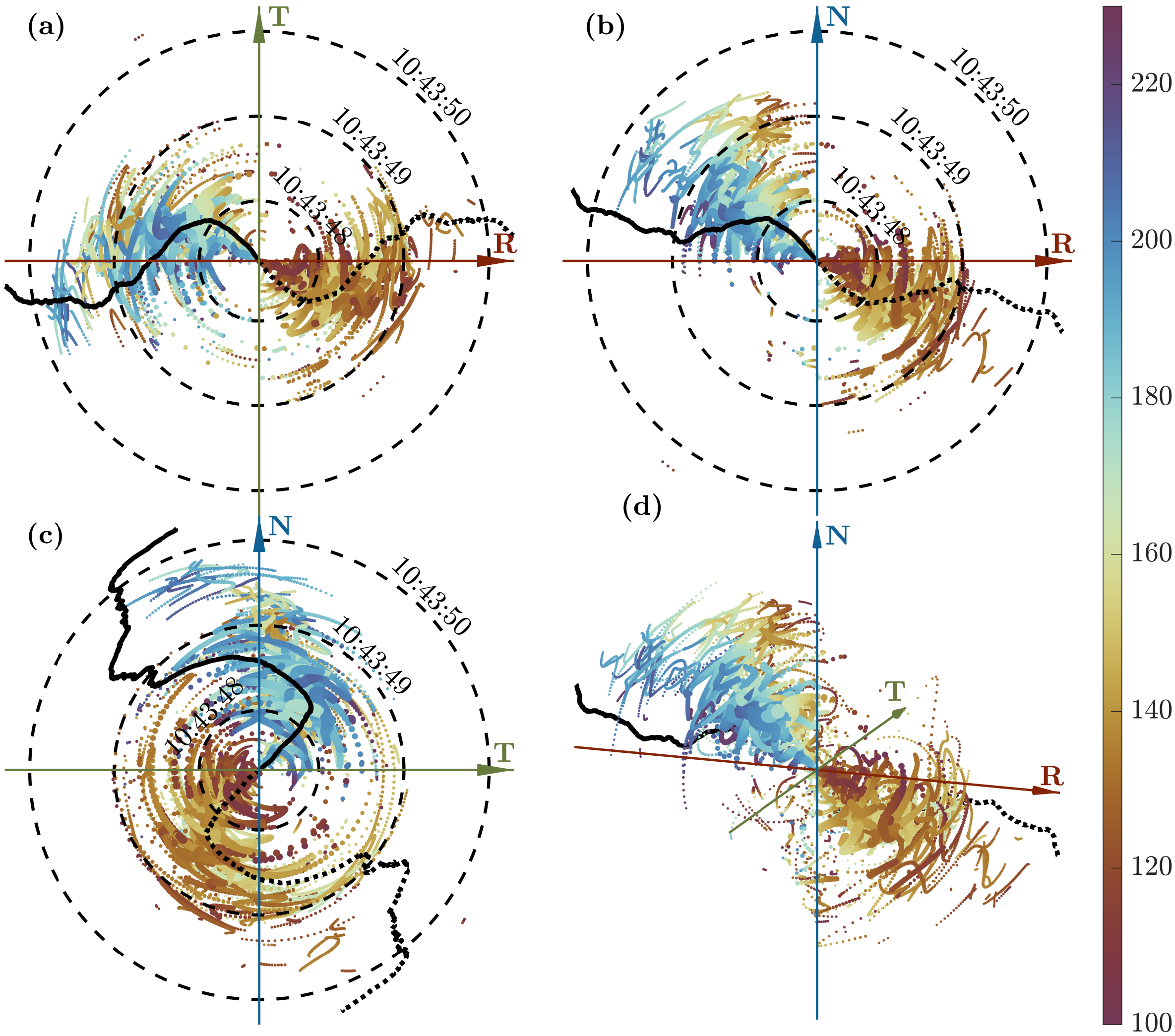

**Figure 4.** Distribution of unit wave vectors for the burst waveform starting at 10:43:47.29 UTC on November 3, 2018. In each of the plots, the distance from the origin is time, the size of the dot is the wave spectral density, and the color is the frequency in the plasma frame, $f\prime$. The solid black curve indicates the direction of the magnetic field and the dashed black curve is the anti-parallel vector to the background magnetic field direction. (a) RT-plane spherical projection. (b) RN-plane spherical projection. (c) TN-plane spherical projection. (d) Three-dimensional distribution in RTN coordinates.

In the spacecraft frame, the frequencies of the sunward waves are less than the anti-sunward waves, however, there is a significant Doppler shift associated with both propagation directions due to the solar wind bulk velocity. The sunward propagating waves will have experienced a Doppler down-shift and thus have a higher frequency in the plasma frame; oppositely, the anti-sunward waves will have been Doppler up-shifted in the spacecraft frame and will have a lower frequency in the plasma frame. The frequency of the whistler waves in the plasma frame, $f\prime$, can be mapped from the spacecraft frame frequency, $f$, using the cold plasma whistler dispersion, plasma density $n$, background magnetic field $B$ from the MAG, solar wind bulk velocity $v_{SW}$, wave normal angle $\theta_{\mathbf{k}\cdot\mathbf{B}}$ with respect to the background magnetic field, and the angle between the wave vector and velocity. A detailed description and illustration of this procedure are provided in Agapitov et al. (2020). The same frequency domain panels in the panels for Fig. 2(e-h) are shown with the corrected (recalculated to the solar wind plasma frame) anti-sunward and sunward Doppler shifts in Fig. 3(a-d) and Fig. 3(e-h), respectively. The anti-sunward waves cover a frequency range of 100 Hz to 180 Hz ($0.065 — 0.12 f_{\mathrm{ce}}$) and the sunward waves range from 150 Hz to 225 Hz ($0.1 — 0.15 f_{\mathrm{ce}}$). The asymmetry in the counter-propagating frequencies is an interesting outcome regarding the possible wave generation mechanism and will be discussed further in Section 4.

The evolution of the distribution of wave normal $\vec{k}$ relative to the background magnetic field direction of the counter-propagating whistlers is shown in Fig. 4. Each point represents the wave's normal direction with the unit vector $\vec{n} = \vec{k}/|k|$, of a single frequency and time from the Doppler-shifted spectrogram. The size of the point represents the wave spectral density, the color is the frequency in the plasma frame, and the radius is time; i.e. the distribution of unit wave vectors at each time are shown as concentric spheres. Panels (a-c) show the spherical projections of the 3D distribution in (d) onto 2D planes with the dashed circles indicating the times corresponding to the major x-axis ticks in Fig. 3. The solid black curve indicates the direction of the magnetic field while the dashed black curve is the anti-parallel vector to the background magnetic field. The counter-propagating whistler waves are observed to propagate in cones surrounding the magnetic field and the wave amplitude is largest during the magnetic rotation. The anti-sunward waves maintain quasi-parallel propagation throughout the burst at all frequencies. The lower frequency sunward waves experience more oblique propagation along the N-component as the magnetic field rotates at the end of the interval, however, the wave vector direction remains largely unchanged.

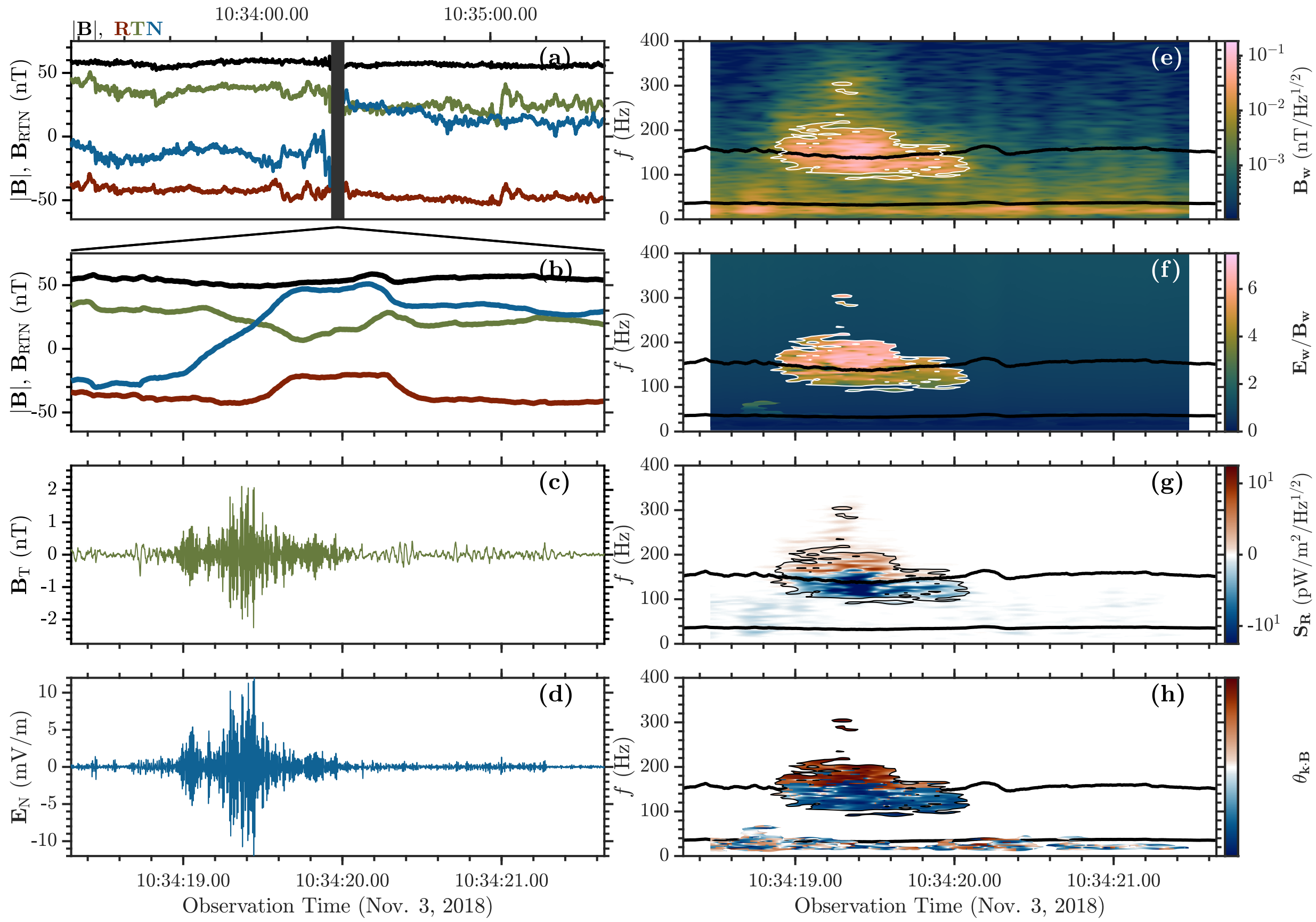

**Figure 5.** The burst interval staring at 10:34:18.22 UTC on November 3, 2018. (a) Dynamics of the magnetic field in RTN coordinates from the MAG over a short time interval around the burst. (b) The shaded region of (a) shows the magnetic field only during the burst. (c) Burst waveform of the magnetic field T-component, $\delta B_{\mathrm{T}}$. (d) Burst waveform of the electric field N-component, $\delta E_{\mathrm{N}}$. (e) Spectrogram of the magnetic field burst waveforms (the white contour indicates the spectral power threshold applied to the data in panels (f-h)). (f) The ratio of electric spectrogram to magnetic spectrogram; the white contour denotes the data boundary with lower amplitude regions omitted for clarity and the background coloring represents the expected $E_{\mathrm{w}}/B_{\mathrm{w}}$ ratio for parallel propagating whistler waves. (g) The radial component of the Poynting vector, $S_R$, is calculated from the spectrograms of each component and transformed to a symmetric log scale as discussed in the text. (h) Wave normal angle with respect to the background magnetic field, $\theta_{\mathbf{k}\cdot\mathbf{B}}$, ranging from 0° to 180° and indicating parallel and anti-parallel propagation, respectively. The lower and upper solid black curves in (e)-(h) indicate $f_{\mathrm{lh}}$ and $0.1 f_{\mathrm{ce}}$, respectively.

The case at 10:34:18.22 UTC is another example of counter-propagating waves and is presented in Fig. 5 in the same format as Fig. 2. During this interval, there is a rapid rotation of the N-component of the magnetic field and an associated magnetic dip (approximately 10% variation) collocated with the whistler wave burst. The wave burst amplitude is maximal at the steepest part of the gradient of $\mathbf{B}_{\mathrm{N}}$ from the MAG. For the $E_{\mathrm{w}}/B_{\mathrm{w}}$ ratio in Fig. 5(f) there is an increasing gradient with frequency for the measured ratio, which is again much larger than the expectation. The Poynting and (WNA) in Fig. 5(g) and Fig. 5(h), respectively, reveal the counter-propagating nature of the burst with quasi-parallel propagation of both the sunward and anti-sunward waves. In the spacecraft frame, the anti-sunward waves occupy a frequency range of approximately 160 Hz to 225 Hz ($0.11$ — $0.15 f_{\mathrm{ce}}$), but a Doppler correction (not shown) reveals the anti-sunward waves experience a Doppler upshift in the spacecraft frame from approximately 110 Hz to 180 Hz ($0.07$ — $0.12 f_{\mathrm{ce}}$) in the plasma frame. Similarly, the sunward waves are Doppler down-shifted from approximately 150 Hz to 240 Hz ($0.1$ — $0.16 f_{\mathrm{ce}}$) in the plasma frame to 90 Hz to 160 Hz ($0.06$ — $0.11 f_{\mathrm{ce}}$) in the spacecraft frame.

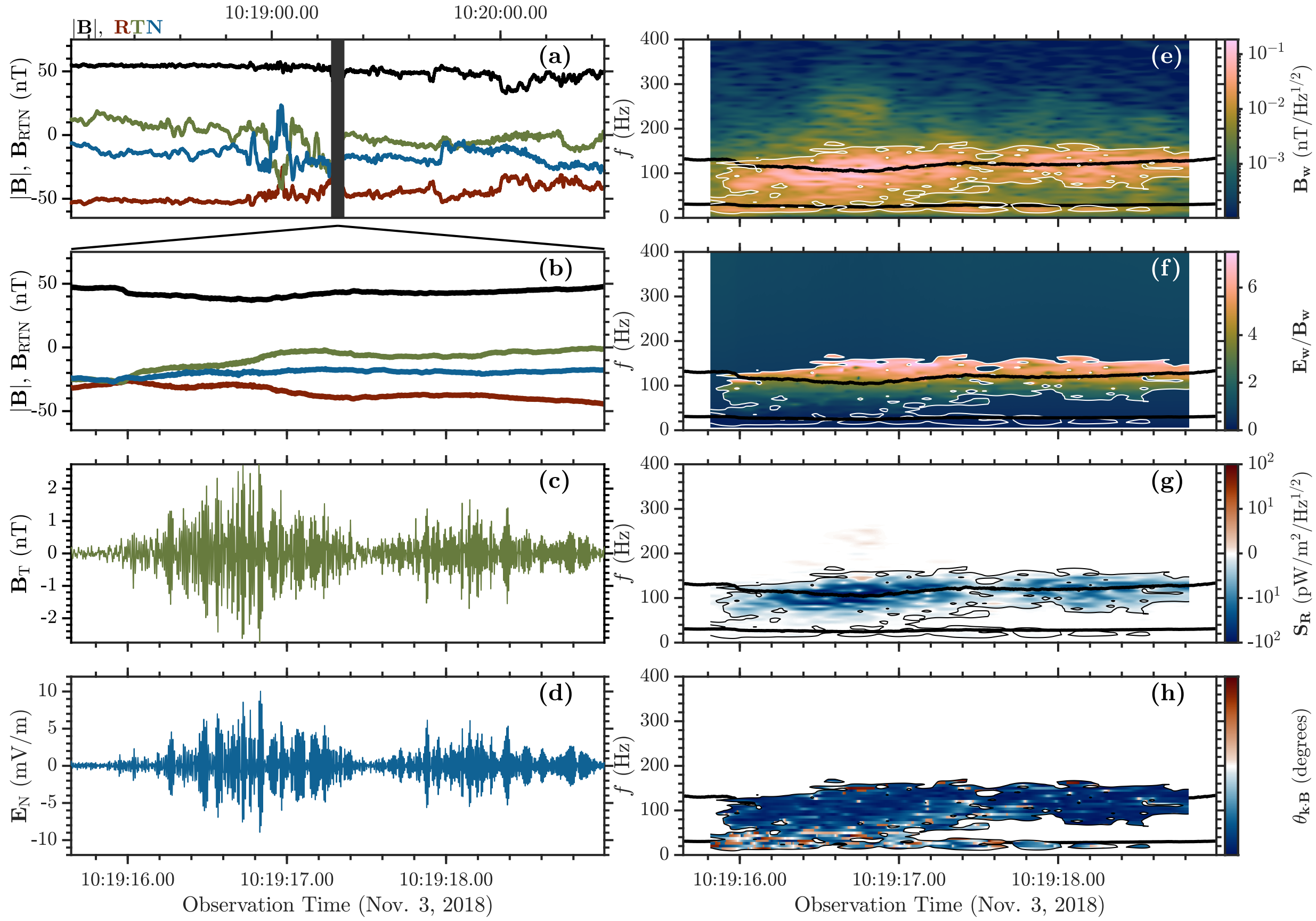

**Figure 6.** The burst interval staring at 10:19:15.56 UTC on November 3, 2018. (a) Dynamics of the magnetic field in RTN coordinates from the MAG over a short time interval around the burst. (b) The shaded region of (a) shows the magnetic field only during the burst. (c) Burst waveform of the magnetic field T-component, $\delta B_{\mathrm{T}}$. (d) Burst waveform of the electric field N-component, $\delta E_{\mathrm{N}}$. (e) Spectrogram of the magnetic field burst waveforms (the white contour indicates the spectral power threshold applied to the data in panels (f-h)). (f) The ratio of electric spectrogram to magnetic spectrogram; the white contour denotes the data boundary with lower amplitude regions omitted for clarity and the background coloring represents the expected $E_{\mathrm{w}}/B_{\mathrm{w}}$ ratio for parallel propagating whistler waves. (g) The radial component of the Poynting vector, $S_R$, is calculated from the spectrograms of each component and transformed to a symmetric log scale as discussed in the text. (h) Wave normal angle with respect to the background magnetic field, $\theta_{\mathbf{k}\cdot\mathbf{B}}$, ranging from 0° to 180° and indicating parallel and anti-parallel propagation, respectively. The lower and upper solid black curves in (e)-(h) indicate $f_{\mathrm{lh}}$ and $0.1 f_{\mathrm{ce}}$, respectively.

Fig. 6 presents the burst at 10:19:15.56 UTC in the same format as Fig. 2. The burst interval is situated in a narrow magnetic well with an approximately 30% decrease of the background magnetic field magnitude at the well minimum that occurs at the end of a period of MHD scale magnetic fluctuations. The whistler wave fluctuations are broken into two clear wave packets that are maximal on the leading and trailing edges of the well. The Poynting flux and WNA reveal that the whistler waves are nearly all propagating sunward. The $E_{\mathrm{w}}/B_{\mathrm{w}}$ ratio is greater than the theoretical expectation shown as the background in Fig 6f, but we note the striking gradient in frequency that will be used in Sec. 3 to empirically calibrate the EFI antennas. In the spacecraft frame, the whistlers cover a frequency range of 60 Hz to 160 Hz ($0.06$ — $0.16 f_{\mathrm{ce}}$) and are Doppler

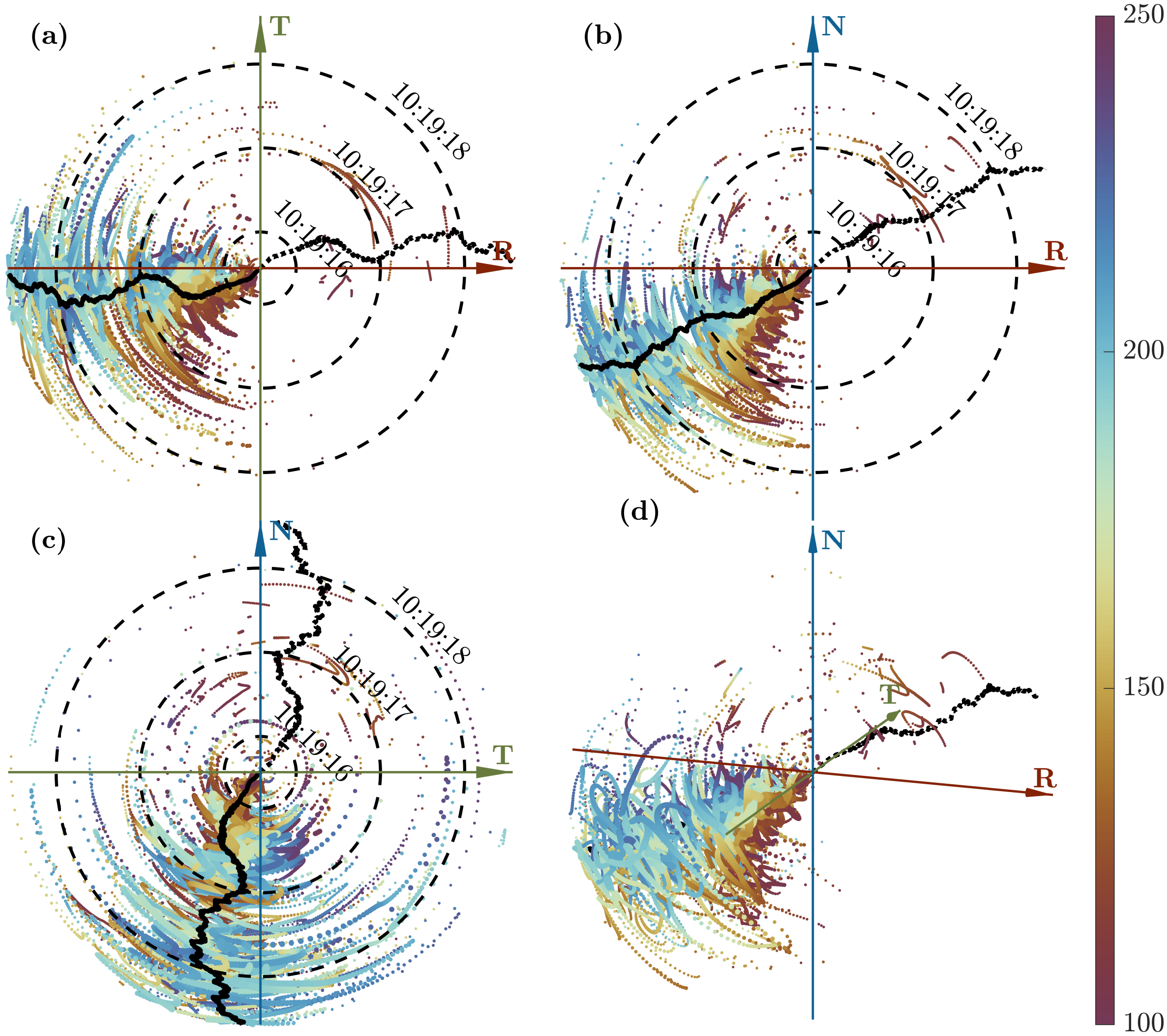

**Figure 7.** Distribution of unit wave vectors for the burst waveform starting at 10:19:15.56 UTC on November 3, 2018. In each of the plots, the distance from the origin is time, the size of the dot is the wave spectral density, and the color is the frequency in the plasma frame, $f\prime$. The solid black curve indicates the direction of the magnetic field and the dashed black curve is the anti-parallel vector to the magnetic field. (a) RT-plane spherical projection. (b) RN-plane spherical projection. (c) TN-plane spherical projection. (d) Three-dimensional distribution in RTN coordinates.

down-shifted from a plasma frame frequency of 100 Hz to 250 Hz ($0.1 — 0.25 f_{\rm ce}$).

In Fig. 7 we present the unit wave vector direction evolution for this sunward propagating burst at 10:19:15:56 UTC. The 3D distribution in Fig. 7(d) shows the propagation of the wave burst is quasi-parallel along the magnetic field directed towards the Sun. In the RT-plane and RN-plane the wave vectors are highly collimated along the magnetic field; there is also a weak dependence on the frequency where the lowest frequency waves are more oblique. In the TN-plane the distribution is more widely dispersed around the parallel field and has very little frequency dependence; this more scattered distribution can largely be attributed to the dominance of the R-component of the background magnetic field seen in Fig. 6(b) that is not captured in Fig. 7(c).

Another example of exclusively sunward propagation is given by the case at 10:30:30.15 UTC shown in Fig. 8. A rotation of the magnetic field and corresponding dip in the magnitude of the magnetic field by nearly 50% is collocated with a very short burst of whistler waves. The magnetic well observed here has more of a form of a step down in magnitude that takes much longer to recover. It is observed that the whistler wave burst is localized to the initial sharp gradient, similar to the gradient localization in the previous case, but this time only on the leading edge as the trailing edge is a gradual recovery. The Poynting flux reveals that this wave packet contains only sunward propagating waves; simi-lar to the other cases, the waves are again nearly parallel to the background magnetic field as shown in the WNA plot. This case has a lower frequency than the rest of the cases and the $E_{\mathrm{w}}/B_{\mathrm{w}}$ ratio at low frequencies is observed to approach the ratio expected from a cold plasma dispersion (see Fig. 8f). This indicates that the effective antenna length is close to the geometrical antenna length at these lower frequencies; this is consistent with observations reported by Mozer et al. (2020).

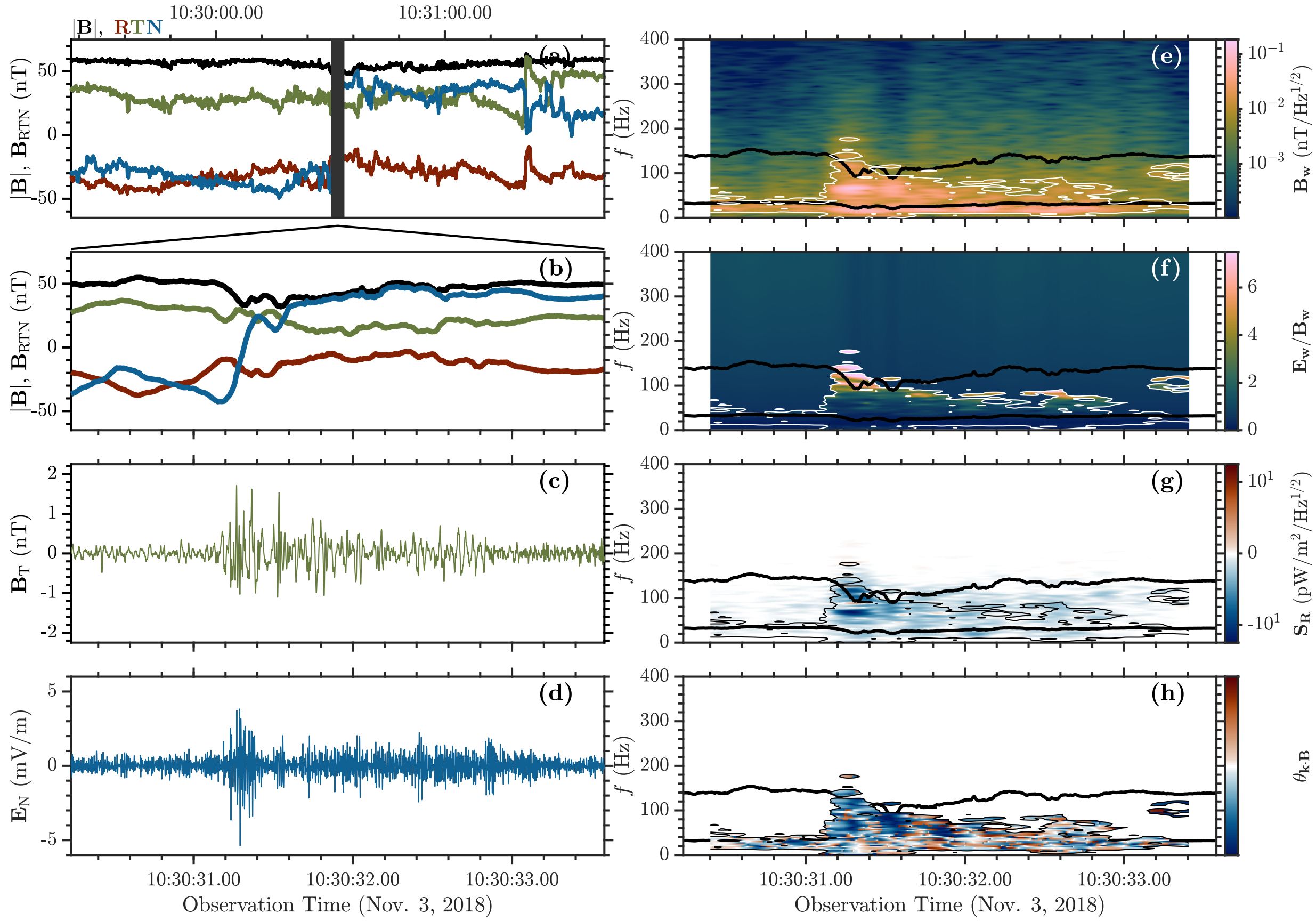

**Figure 8.** Analysis of the burst waveform starting at 10:30:30.15 UTC on November 3, 2018. (a) Dynamics of the magnetic field in RTN coordinates from the MAG over a short time interval around the burst. (b) The shaded region of (a) shows the magnetic field only during the burst. (c) Burst waveform of the magnetic field T-component, $\delta B_{\mathrm{T}}$. (d) Burst waveform of the electric field N-component, $\delta E_{\mathrm{N}}$. (e) Spectrogram of the magnetic field burst waveforms. (f) The ratio of electric spectrogram to magnetic spectrogram; the white contour denotes the data boundary with lower amplitude regions omitted for clarity and the background coloring represents the expected $E_{\mathrm{w}}/B_{\mathrm{w}}$ ratio for parallel propagating whistler waves. (g) The radial component of the Poynting vector is calculated from the spectrograms of each component and transformed to a symmetric log scale as discussed in the text. (h) Wave normal angle with respect to the background magnetic field, $\theta_{\mathbf{k}\cdot\mathbf{B}}$, ranging from 0° to 180° and indicating parallel and anti-parallel propagation, respectively. The lower and upper solid black curves in (e)-(h) indicate $f_{\mathrm{lh}}$ and $0.1 f_{\mathrm{ce}}$, respectively.

The case at 10:33:31.03 is presented in Fig. 9; in this case, we observe the waves in a calm magnetic field during the DBM event. The Poynting flux shows that the packets of whistler waves are all propagating primarily away from the Sun and the WNA shows they are nearly parallel to the background magnetic field. Given that anti-sunward waves do not strongly interact with the solar wind particle populations, and that there is a lack of a strong inhomogeneity in the magnetic field, it is possible that the waves observed here have traveled from another source to the spacecraft; this interpretation is supported by the lack of sunward waves collocated with this event and the lack of a magnetic inhomogeneity to indicate some source mechanism. The amplitude of these waves is also significantly less than the other four cases, this is further evidence they may have originated elsewhere in the solar wind and experienced some light damping as they have propagated to the location of PSP. The hatched regions in the frequency-dependent plots serve only to remove signals introduced by voltage perturbations due to plasma clouds generated by dust impacts on the spacecraft (Malaspina et al. 2020; Page et al. 2020). These waves are again Doppler shifted to a frequency of 160 Hz yo 220 Hz ($0.1 — 0.14 f_{\mathrm{ce}}$) in the spacecraft frame from a frequency of 100 Hz to 160 Hz ($0.06$ — $0.1 f_{\mathrm{ce}}$) in the plasma frame.

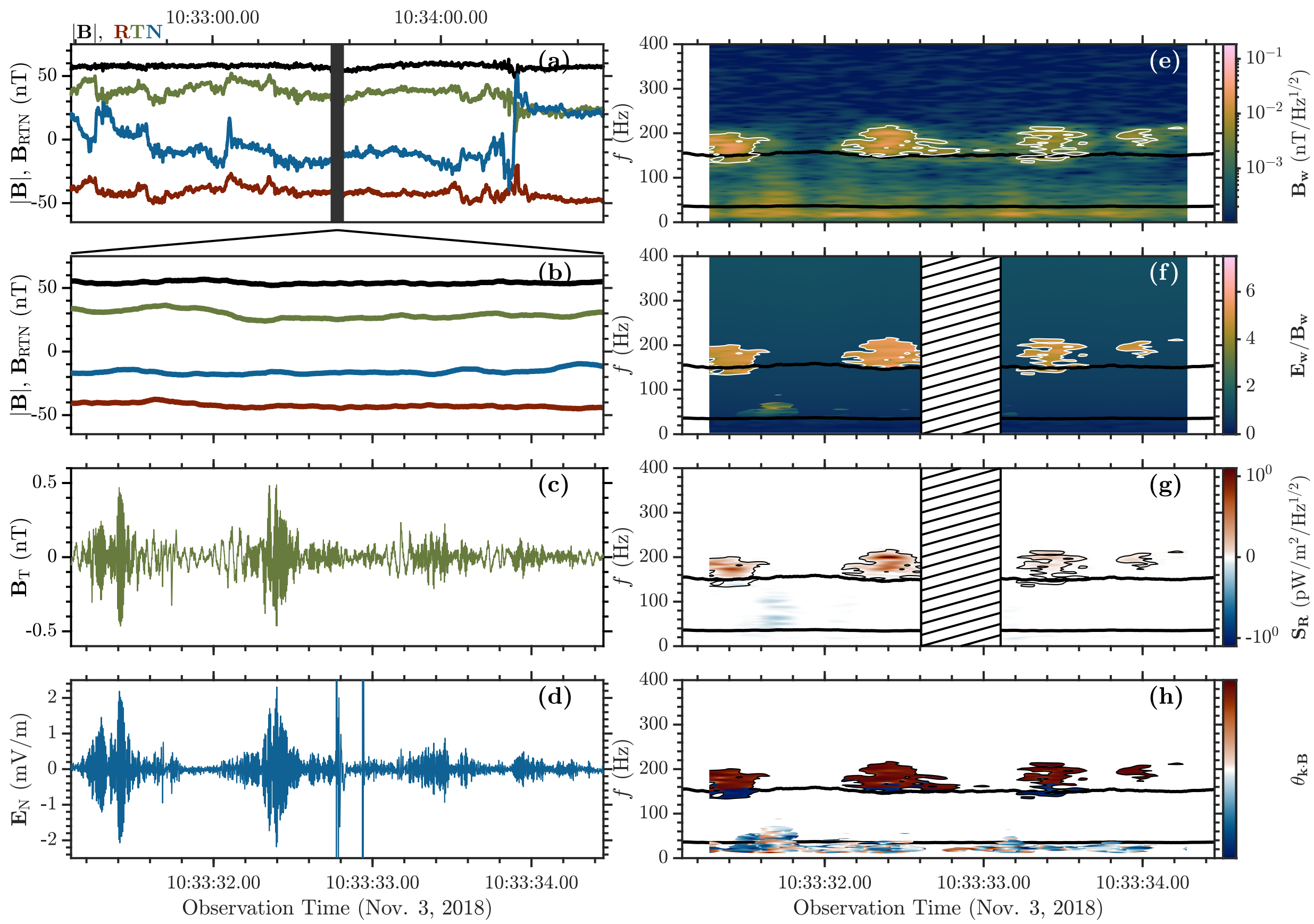

**Figure 9.** The burst interval staring at 10:33:31.03 UTC on November 3, 2018. (a) Dynamics of the magnetic field in RTN coordinates from the MAG over a short time interval around the burst. (b) The shaded region of (a) shows the magnetic field only during the burst. (c) Burst waveform of the magnetic field T-component, $\delta B_{\mathrm{T}}$. (d) Burst waveform of the electric field N-component, $\delta E_{\mathrm{N}}$. (e) Spectrogram of the magnetic field burst waveforms (the white contour indicates the spectral power threshold applied to the data in panels (f-h)). (f) The ratio of electric spectrogram to magnetic spectrogram; the white contour denotes the data boundary with lower amplitude regions omitted for clarity and the background coloring represents the expected $E_{\mathrm{w}}/B_{\mathrm{w}}$ ratio for parallel propagating whistler waves. (g) The radial component of the Poynting vector, $S_R$, is calculated from the spectrograms of each component and transformed to a symmetric log scale as discussed in the text. (h) Wave normal angle with respect to the background magnetic field, $\theta_{\mathbf{k}\cdot\mathbf{B}}$, ranging from 0° to 180° and indicating parallel and anti-parallel propagation, respectively. The lower and upper solid black curves in (e)-(h) indicate $f_{\mathrm{lh}}$ and $0.1 f_{\mathrm{ce}}$, respectively.

## 3. APPLICATION TO THE FIELDS INSTRUMENT

The five cases presented in the previous section, and specifically the two counter-propagating cases, offer an opportunity to calibrate the electric field antennas on the FIELDS instrument in the whistler waves frequency range. The methodology presented here is similar to that used for the wave cases presented by Mozer et al. (2020). There are two frames of interest in this problem, the spacecraft frame (denoted by unprimed quantities) and the plasma frame (denoted by primed quantities), with the plasma frame moving at a relative velocity, **v**, with respect to the spacecraft frame. The electromagnetic fields in the spacecraft frame are related to the plasma frame fields by the Lorentz transformations for special relativity (Feynman 1964),

$$E_{\parallel} = E'_{\parallel} \tag{2}$$

$$B_{\parallel} = B'_{\parallel} \tag{3}$$

$$E_{\perp} = \frac{\left(\vec{E}' - \vec{v} \times \vec{B}'\right)_{\perp}}{\sqrt{1 - v^2/c^2}} \tag{4}$$

$$B_{\perp} = \frac{\left(\vec{B}' + \frac{\vec{v} \times \vec{E}'}{c^2}\right)_{\perp}}{\sqrt{1 - v^2/c^2}} \tag{5}$$

where here the transformation is in the format of the so-called "inverse" transformation due to our choice of the spacecraft frame as the unprimed frame. In the case where $v << c$, as is the case for the PSP observations

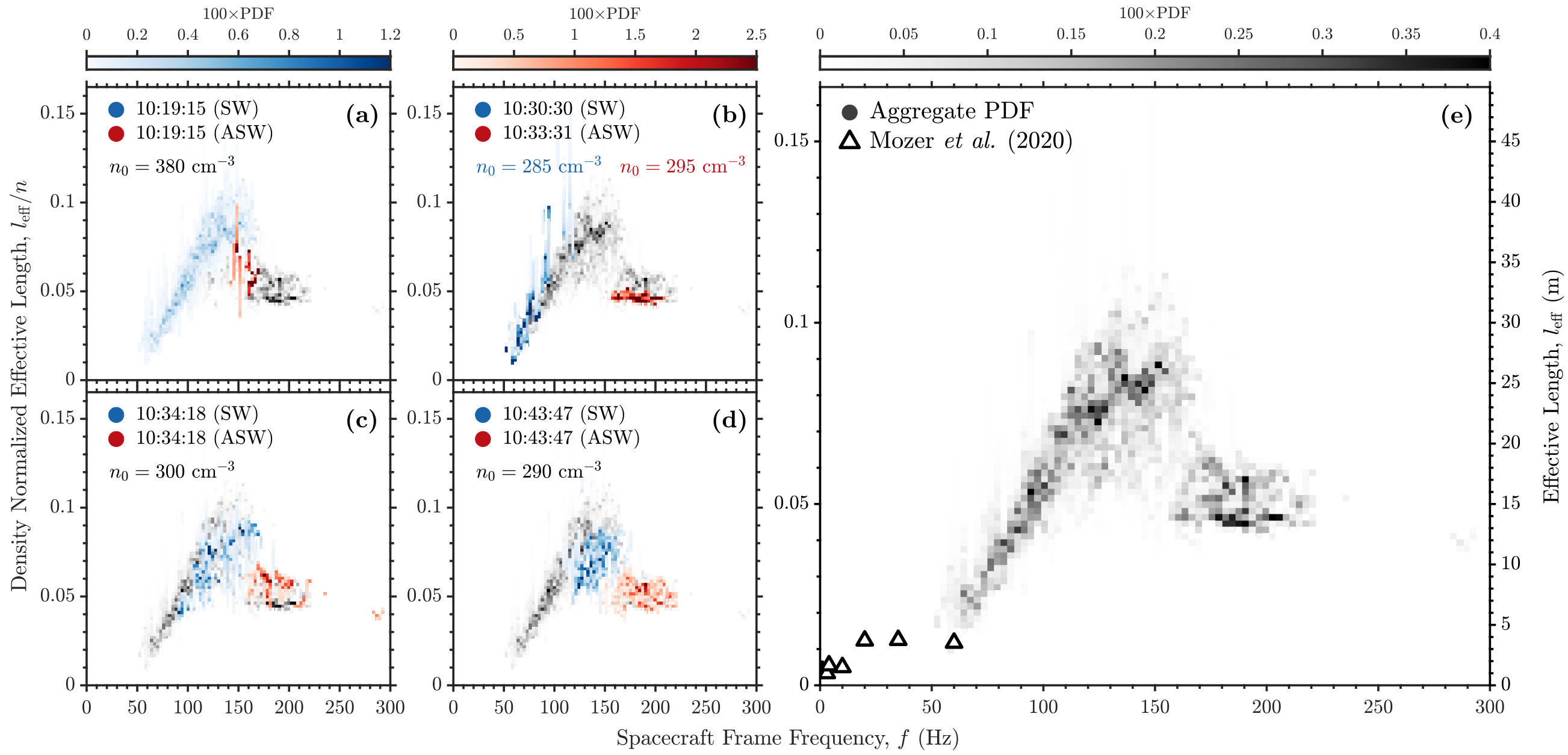

**Figure 10.** (a)-(e) The bivariate PDFs of the density normalized effective antenna length and wave frequency for the five cases of whistler bursts in the one-hour interval presented herein: (a) sunward case from Fig. 6; (b) sunward case at 10:30:30.15 and anti-sunward case from Fig. 9; (c) counter-propagating case at 10:34:18.22; (d) counter-propagating case from Fig. 2; and (e) the aggregate of all the cases. In (a) through (d) the sunward waves are shown by blue markers and anti-sunward by red markers, with the aggregate from (e) shown in the background for clarity. The hollow black triangles in (e) represent the result from Mozer et al. (2020) for comparison. The right axis is an estimate of the effective length in meters using the density of $n = 300$ cm$^{-3}$ from Mozer et al. (2020) and is similar to the density for the five cases presented here.

presented herein, Eqs. (4) and (5) are reduced to,

$$E_\perp = \left(\vec{E}' - \vec{v} \times \vec{B}'\right)_\perp \tag{6}$$

$$B_\perp = B'_\perp \tag{7}$$

The effect of special relativity on the whistler wave measurements is an additional electric field proportional to $\mathbf{v} \times \mathbf{B'}_\text{w}$ associated with the whistler wave that is measured by the FIELDS electric antennas, where $\mathbf{B}'_\text{w}$ refers to the magnetic field waveform in the plasma frame. It is well documented that electric field antenna response is further modified by an undetermined transfer function due to coupling between the spacecraft and plasma (Gurnett 1998). This coupling can be determined by the spacecraft geometry, frequency and wavelength of the electromagnetic or electrostatic fields being measured, and the plasma parameters. A simple way to correct for this interaction is to use an effective length, $l_{\text{eff}}$, calibration (Mozer et al. 2020; Hartley et al. 2015, 2016),

$$\frac{l_{\text{ant}}}{l_{\text{eff}}} \vec{E}_{\text{exp}} = \vec{E}'_\text{w} - \mathbf{v} \times \vec{B}'_\text{w} \tag{8}$$

where $\vec{E}_{\text{exp}}$ represents the measured electric field using the nominal antenna length, $l_{\text{ant}} = 3.5$ m. The effective antenna length for a given frequency can be estimated from whistler waves measured by PSP using the apparent velocity from the ratio of the measured $E_\text{w}$ and $B_\text{w}$, the plasma flow speed $\vec{V}_{SW}$, the unit wave vector $\vec{k}/|k|$, and the expected whistler phase velocity from the cold plasma dispersion for the frequency in the plasma frame, $v_{\text{ph}}$,

$$l_{\text{eff}}(\omega) = l_{\text{ant}} \frac{|\vec{E}_\text{w}(\omega)|/|\vec{B}_\text{w}(\omega)|}{v_{\text{ph}}(\omega') + \vec{V}_{SW} \cdot \hat{\vec{k}}(\omega)} \tag{9}$$

where $|\vec{E}_\text{w}(\omega)|$ denotes the amplitude of a wave at a specific frequency in the spacecraft frame; $\omega$ is the wave frequency in the spacecraft frame; $\omega'$ is the wave frequency in the plasma frame.

Using Eq. (9) and the Doppler corrected wave spectra in the plasma frame, like those presented in Fig. 3, we estimate the effective length $l_{\text{eff}}$ for each of the five DBM cases. Fig. 10 displays the result of this analysis with $l_{\text{eff}}$ normalized to the plasma density. Panels (a) through (d) present the bivariate PDF of the results for each of the individual cases overplotted on the aggregate of all the cases with sunward and anti-sunward waves indicated by blue and red, respectively. The overlap of the common frequencies in both the counter-propagating cases represents a strong validation of the result as the Lorentz-transformed electric field and Doppler shift are opposite for each wave propagation direction. Fig. 10(e) displays the aggregate of all the cases along with a comparison of the result from Mozer et al. (2020) with frequencies of 1, 3, 4, 10, 20, 35, and 60 Hz. There is excellent agreement among all five cases and with the Mozer et al. (2020) results at low frequencies. The right axis shows an estimate of the $l_{\rm eff}$. It is observed that at low frequencies up to about 60 Hz, $l_{\rm eff}$ is less than or equal to the antenna geometrical length of 3.5 m. Above 60 Hz $l_{\rm eff}$ increases to 22-28, i.e. nearly a factor 6-8 greater than expected, and peaks locally at a frequency of $140-150$ Hz before decreasing and appearing to have a plateau at and above 160-170 Hz (over to the frequency range investigated) at around $13-17$ m. Overall, $l_{\rm eff}(\omega)$ ranges from ∼5 m to ∼28 m for the frequency range 35 Hz to 300 Hz. The obtained $l_{\rm eff}$ has a comparably large spread of values (10-15%), which is a result of the measurements regime and approximations used for processing: the PSP plasma data sampling rate of 1 sec is a result of plasma parameters averaging during the processing intervals when the actual solar wind bulk velocity and plasma density can vary in the range of 5-10% (Mozer et al. 2023) affecting the evaluated wave properties; the approximation of field-aligned whistler waves propagation (we filtered out waves with WNA above 20 degrees for the estimations) can lead to less than 5% error of estimation since the maximum of WNA distribution is at $\sim 5$ degrees; the possible dependence on the background plasma density (reported by Hartley et al. (2017) for the Van Allen Probes measurements). The slightly higher values of $l_{\rm eff}$ obtained at higher density (Figure 10a) will need further investigations. Such a large range of effective lengths over a small frequency range of the FIELDS EFI instrument (which covers DC to 60 kHz) is concerning regarding the reliability and potential for misinterpretation of electric field measurements made by PSP but is not wholly unexpected. The effective length $l_{\rm eff}$ of the electric field sensor for the spacecraft measurements in very variable plasma conditions cannot be accurately obtained in the ground test (Imachi et al. 2007; Hartley et al. 2017). Thus the in-flight calibrations, which can be based on the suitable plasma models and the properties of plasma wave modes, can be applied. The wide range of plasma density in the solar wind (or magnetosphere plasma (Hartley et al. 2017)) can lead to different approximations for the antenna effective length, for example, when the geometrical antenna scale is less about the local Debye length in plasma, and or the antennas lengths are of the same order of the spacecraft size. This case is regular for the PSP EFI measurements in the solar wind with Debye length of about units or tens of m and the relatively short antennas (3.5 m), which are less about the PSP linear sizes. We used here the waves processed in the cold plasma approximation to estimate the frequency dependence of $l_{\rm eff}(\omega)$ for the cases collected during the one-hour interval with similar plasma conditions. The overestimation of the electric field from the measurements on board the two Van Allen Probes in the Earth magnetosphere with the short boom antenna (dependent on the background plasma density and wave frequency) was reported by Hartley et al. (2017). Hartley et al. (2017) discussed a possible source of this anomalous gain to be in the geometry of the voltage-biased surfaces around the sensors. The extra gain was significant for the shorter spin axis sensors (the length was 14 m versus 40 and 50 m for the spin plane sensors) - the short boom axial antennas provide 1.5-3 times overestimated electric field measurements (Hartley et al. 2017). We presume that the geometry of the voltage-biased surfaces of the PSP spacecraft can be a source of the observed overestimation of electric field measurements in the frequency range from 50 to 160 Hz observed in plasma with a density of 300-380 cm-3. The estimates presented here could be used in conjunction with those made by Mozer et al. (2020) to assist with the interpretation of the FIELDS antennas and guide future attempts at empirical calibration. It should be noted that these results are for cases with similar densities and magnetic field strengths and caution should be used in applying these effective lengths to other regimes.

## 4. DISCUSSION

High-sampling-rate SCM and EFI PSP observations of electromagnetic fields in the burst mode (150 kS/s) have revealed the occurrence of localized bursts of low-frequency whistler wave packets: the typical frequency range is 100 Hz to 300 Hz ($0.05 — 0.2 f_{\rm ce}$) in the spacecraft frame (in an agreement with statistical studies by Jagarlamudi et al. (2021); Cattell et al. (2022)). These waves often coincide with local minima of the magnetic field magnitude or with edges of magnetic switchbacks containing sharp rotations of the background magnetic field direction: the distribution of the whistler waves amplitudes in their relation to the magnetic field magnitude perturbations ($\Delta|B|$) and magnetic field perturbation ($|\Delta\vec{B}|$) is presented in Fig. 11 indicating collocation of wave sources with magnetic dips for highest amplitudes of sunward propagating waves (Fig. 11(a)), and collocation of wave sources (the counter-propagating waves) with changes of magnetic field direction (Fig. 11(b)). The high amplitude predominantly sunward waves appear to be strongly associated with the local dips of the background magnetic field (high $\Delta|B|$ in Fig. 11a). The counter-streaming cases were observed simultaneously with sharp changes in magnetic field direction ($|\Delta\vec{B}|$) but do not contain a dip of background magnetic field $|B|$. The anti-sunward case was found to be unrelated to any inhomogeneity in the magnetic field (we presume that the waves observed with lower values of $|\Delta\vec{B}|$ are anti-sunward waves). The sunward and counter-streaming cases observations suggest that we observe them in the vicinity of the source region – the sunward whistlers can interact efficiently with the strahl, so are rapidly decaying interacting with the strahl electrons in the source region (possibly by the gyrosurfing mechanism (Artemyev et al. 2013; Kis et al. 2013)). This presumably determines the high correlation of sunward whistlers with magnetic field inhomogeneities. However, anti-sunward waves, even generated in the vicinity of magnetic field inhomogeneities, can propagate long distances without s significant decay. Wave amplitudes peaked at 2 nT representing up to 0.05 of the background magnetic field magnitude. The polarization of these waves was found to be predominantly quasi-parallel to the background magnetic field confirming the statistics reported by Cattell et al. (2020); Froment et al. (2022). The quasi-parallel propagation appears to be the predominant mode of the observed whistler waves in the solar wind in all the range of heliocentric distances covered by observations by PSP and Solar Orbiter - the probability to observe waves with wave normal angle above 50° is less than 0.01 (Cattell et al. 2022; Kretzschmar, M. et al. 2021; Tong et al. 2019). This makes propagation direction to be the key factor for wave-particle interaction efficiency: the scattering of the strahl electrons by the counter-propagating (sunward propagating) whistler wave is more than an order more efficient than scattering by anti-sunward propagating quasi-parallel waves. All whistler waves observed by Solar Orbiter at $0.5-1.0$ AU were found to propagate from the Sun (Kretzschmar, M. et al. 2021). Thus, the sunward propagating whistler population in the young solar wind (observed by PSP at $25-45R_{\odot}$) can be a potentially key factor for the scattering of the strahl electron population. It raises a question about the mechanism of their local generation process (to explain the heliocentric distances range of generation of these waves reported by Cattell et al. (2022)) and requires a statistical study of whistler wave occurrence rates and their parameters in the young solar wind to include into the consideration both the wave amplitude distribution and wave polarization parameters including their propagation direction.

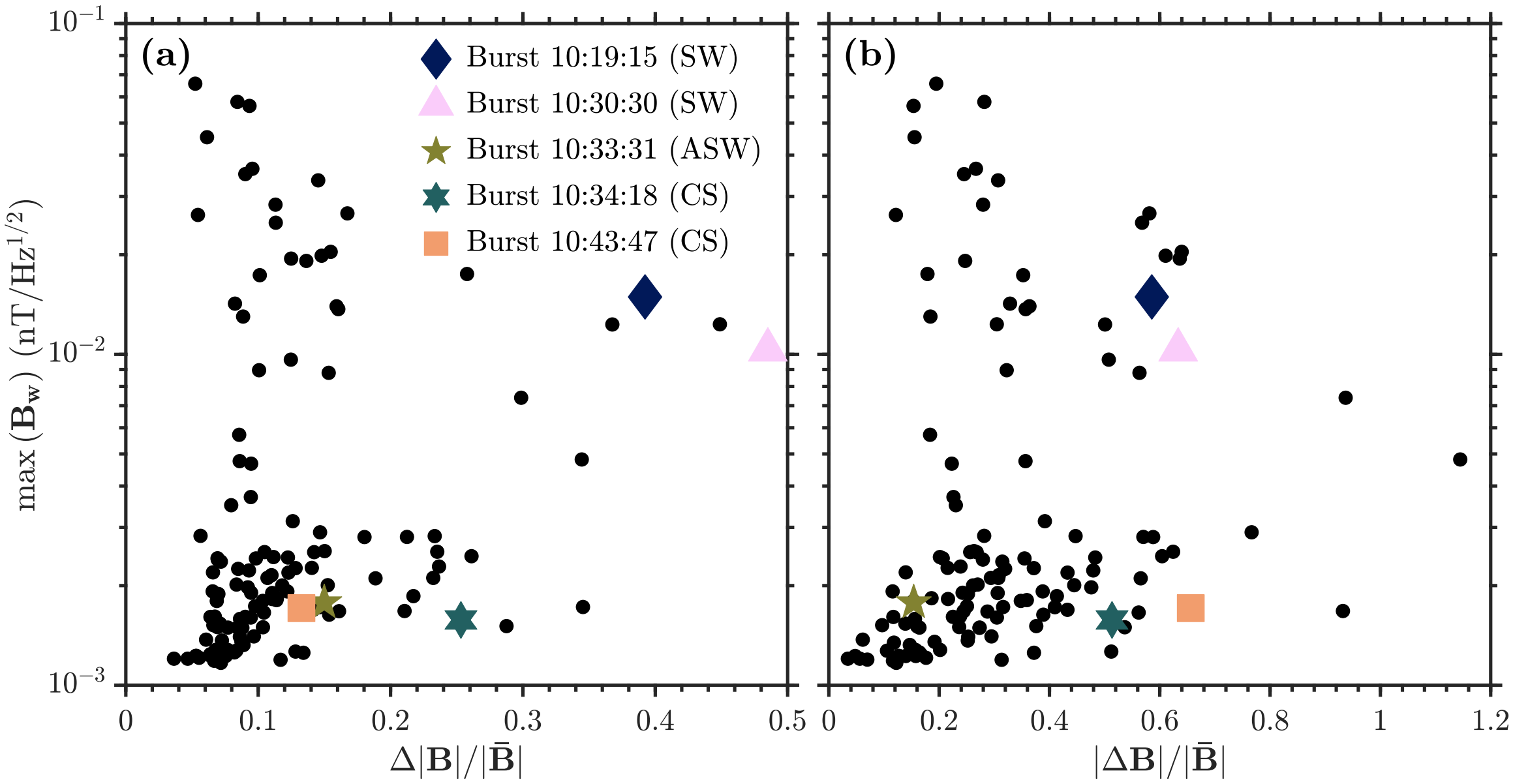

**Figure 11.** Distribution of whistler bursts observed during 10-11 UT (the interval presented in Fig.1) in the wave amplitude and the background magnetic field perturbation domain: (a) On the vertical axis is the maximum amplitude of each temporal bin of the spectral matrices within the frequency range of the whistler waves and on the horizontal axis the normalized perturbation of the magnetic field magnitude of $\Delta|B|/|B|$ during the same time window. (b) The same vertical axis with the horizontal axis represents the perturbation of the magnetic field vector of $|\Delta\vec{B}|/|B|$ during the same time window. The black circles indicate each of the temporal bins while the 5 cases are highlighted by the colored markers in the legend of (a) and are indicated as either sunward (SW), anti-sunward (ASW), or counter-propagating (CS).

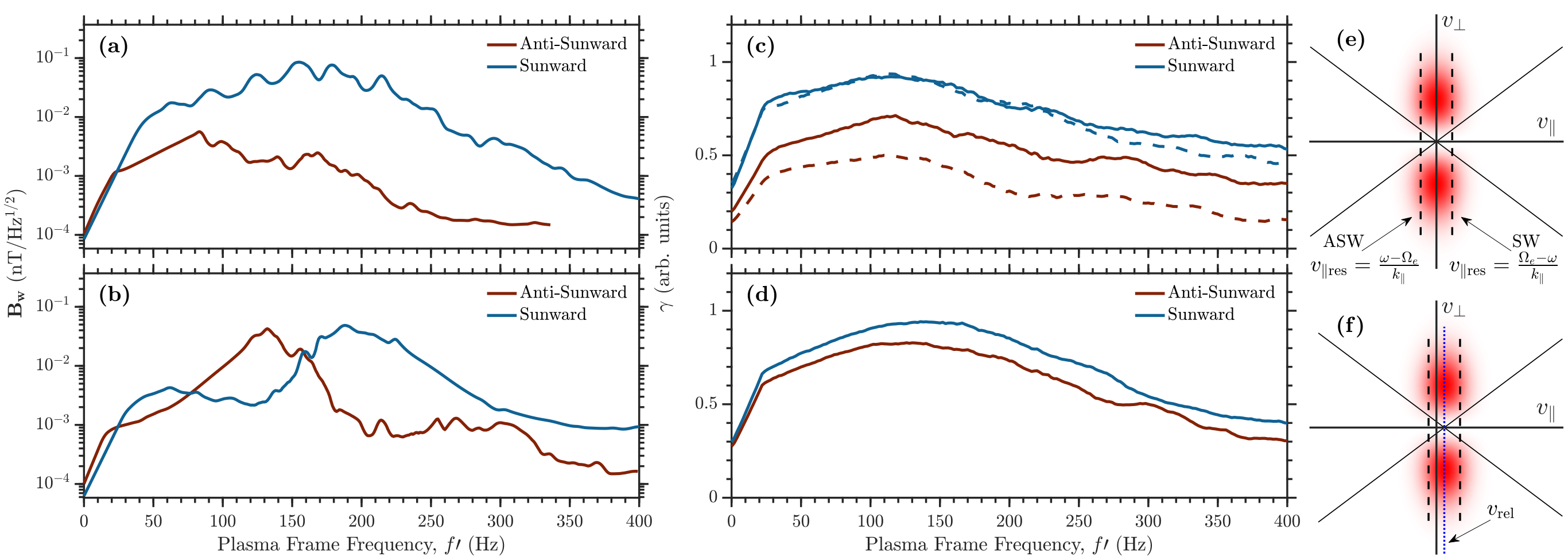

**Figure 12.** (a-d) Comparison of the Doppler shifted spectrum with the growth rates estimate under the quasi-linear approximation. (a) Predominantly sunward waves from the case at 10:19:15 UTC. (b) Counter-propagating waves with shifted frequency peaks from the case at 10:43:47. (c) Predicted growth rates using $T_\perp = 100$ eV and $T_\parallel = 20$ eV, and magnetic hole relative velocity $v_{rel} = v_A$ (solid curves) and $v_{rel} = 2v_A$ (dashed curves). (d) Predicted growth rates using $T_\perp = 100$ eV and $T_\parallel = 20$ eV, and magnetic hole relative velocity $v_{rel} = v_A$. (e) and (f) Schematic descriptions of the wave generation mechanism by trapped particle populations. (e) A symmetric loss cone distribution is equally unstable to whistler waves of the same frequency propagating both sunward (SW) and anti-sunward (ASW). (f) A loss cone distribution with a positive relative drift velocity, $v_{\mathrm{rel}}$, is unstable to counter-propagating whistler waves where the peak frequency of the SW waves is larger than that of the ASW waves; a larger $v_{\mathrm{rel}}$ results in lower-growth rates of the ASW waves.

The lower sampling rate of the electron distribution function evaluated by the SWEAP instrument suite on board PSP (one per 27 s during Encounter 1, and 14 s during later Whittlesey et al. (2020)) doesn't allow direct processing of the electron instabilities inside the discussed magnetic holes (their duration is ∼1 s), so we use the available plasma density measurements and the averaged electron distribution functions for a model for the electron distribution inside the magnetic field minimum. The plasma density didn't have any peculiarities inside the magnetic field magnitude depression region Agapitov et al. (2020). This suggests existing of the hot trapped population with the loose cones and respectively with the transverse anisotropy favorable for the generation of quasi-parallel whistlers. Similar structures with magnetic field magnitude depletion were observed in the vicinity of the Earth bow shock by the four MMS spacecraft with high-time resolution electron measurements aboard (Ahmadi et al. 2018), and these observations nicely confirm our assumptions highlighting the trapped hot electron population. Such distribution (with the proper plasma temperature) can guide the generation of whistler waves propagating along and opposite the background magnetic field through the cyclotron instability (Ahmadi et al. 2018; Tigik et al. 2022; Malaspina et al. 2022; Agapitov O. et al. 2018; Drake et al. 2015). For the magnetic hole, which is in the rest in respect to the background core plasma, i.e. propagating with the bulk solar wind velocity, such a distribution is shown schematically in Fig. 12(e). However, the wave sources are connected with the magnetic field inhomogeneities (like switchback boundaries) propagating in the solar wind plasma with significant relative velocity $\delta V$ in the range of 50 km/s to 150 km/s. This relative velocity leads to the shift of distribution function of the trapped electron population with $\delta V$ (schematically shown in Fig. 12(f)) and corresponding generation of sunward and anti-sunward whistler waves with the difference in the wave frequency (Fig. 12(a,b) present whistler spectra in the plasma frame) and amplitude (Fig. 12(a)). Numerical estimation of the growth rate reveals this difference in Fig. 12c,d: propagation with half of the local Alfvén velocity separates the sunward and anti-sunward waves in frequency with a maximum growth rate at a higher frequency for the sunward wave and at lower for anti-sunward wave (Fig. 12(b)); propagation of magnetic hole with the local Alfvén velocity changes both the frequency and value of the maximum wave growth rate (Fig. 12(c)) – sunward waves have several times higher growth rate than the anti-sunward ones (Fig. 12(d)). Because of the lack of high sampling rate measurements of the electron distribution function, we can provide the qualitative reliance on the observed whistlers properties: the faster propagating magnetic hole can lead to a greater difference in the frequency of growth rates for sunward and anti-sunward waves (Cases in Fig. 2 and Fig. 5 of counter-propagating whistlers in the current paper) up to full suppression of anti-sunward waves generation (Cases in Fig. 6 and Fig. 8 in the current paper and the cases reported by Agapi-tov et al. (2020)). Additionally, observations of sunward whistlers only at their generation regions and observations of anti-sunward propagating whistler waves unrelated to the source regions supports the difference in the wave-particle interaction efficiency: sunward wave are rapidly decaying on the edges of the source region but anti-sunward quasi-parallel waves propagate almost without decay.

## 5. CONCLUSIONS

1. We have identified the presence of intense whistler waves in their generation region from Parker Solar Probe observations at a heliocentric distance of approximately $35.7R_\odot$ and evaluated their parameters:

    - the counter-propagating nature of the sunward and anti-sunward whistler waves indicates that observations were captured in the wave source region;
    - the occurrence of sunward propagating whistler waves is related to local inhomogeneities of the magnetic field often associated with switchback boundaries. Such close relation indicates that sunward propagating whistlers are observed in their generation region and rapidly decay because of efficient wave-particles interaction outside the source;
    - the anti-sunward waves experience slower decay and can be observed far from their source regions;
    - observed waves had mostly quasi-parallel to the background magnetic field wave normal angles (less than 20 degrees). Some deviations from the quasi-parallel propagation (with wave normal angle up to 60 degrees) were found to be connected with the inhomogeneous geometry of the background magnetic field.

2. The generation mechanism for these sunward and anti-sunward propagating quasi-parallel whistler waves is presumably the cyclotron instability (suggested by simultaneous generation of counter-propagating wave bursts) driven by suprathermal electron population (of the order of 100 eV) trapped in magnetic field dips and/or the strahl population with perturbed geometry at local inhomogeneities of the magnetic field (sharp rotations) at switchback boundaries. The derived difference in frequency of sunward and anti-sunward counter-propagating waves confirmed their simultaneous generation in the source region moving in the solar wind frame.

3. The effective length of the PSP antennas applied to electric field measurements being carried out by the Electric Field instrument (EFI) has been processed in the frequency domain of whistler waves making use of the Search Coil Magnetometer (SCM) measurements of magnetic field perturbations and the whistler wave dispersion relation. The simultaneous observations of counter-propagating whistler waves were used to eliminate the moving media effects of solar wind flow. The EFI effective length ($l_{\rm eff}$) is about 1 m for the near DC measurements, grows up to 3.5 m to 4.5 m for frequencies of 10 Hz to 50 Hz (Mozer et al. 2020), and peaks from 4.5 m to ∼25 m when the frequency increases from 50 Hz to 130 − 150 Hz for the plasma density range of 280 − 350 cm-3. Above 150 Hz $l_{\rm eff}$ saturates at approximately 15 m (we covered the frequency range below 300 Hz in this work). From an experimentalist's perspective, perhaps the most consequential outcome of this analysis is that these values of $l_{\rm eff}$ can cause significant overestimation of the measured electric field and require proper calibration before processing the efficiency of wave-particle interactions at frequencies above 50 Hz.

## 6. ACKNOWLEDGMENTS

S.K., O.V.A., H.Y.K were supported by NASA grants 80NNSC19K0848, 80NSSC22K0417, and NSF grant 1914670. O.V.A. was partially supported by NASA grants 80NSSC20K0218, 80NSSC20K0697, and 80NSSC21K1770. The PSP/FIELDS experiment was developed and is operated under NASA contract NNN06AA01C.